\documentclass{elsart3}
\journal{New Astronomy}
\usepackage{natbib}
\usepackage{epsfig}
\usepackage{amssymb}

\begin{document}

\newcommand{\n}     {\noindent}%
\newcommand{\sn}    {\smallskip\noindent}%
\newcommand{\ve}    {\vfill\eject}%
\newcommand{\cge}   {\ensuremath{\,\gtrsim\,}}%
\newcommand{\cle}   {\ensuremath{\,\lesssim\,}}%
\renewcommand{\etal}{\mbox{et al.\ }}%
\newcommand{\eg}    {\mbox{e.g.}}%
\newcommand{\ie}    {\mbox{i.e.}}%
\newcommand{\HST}   {{HST}}%
\newcommand{\iAB}   {{$i'_{AB}$}}%
\newcommand{\zAB}   {{$z'_{AB}$}}%
\newcommand{\Mo}    {{$M_{\odot}$}}%
\newcommand{\Mbh}   {{$M_{bh}$}}%
\newcommand{\arcspt}{\hbox{$.\!\!^{\prime\prime}\!$}}%


\begin{frontmatter}

\title{\huge\sc Did Galaxy Assembly and Supermassive\\
Black-Hole Growth go hand-in-hand?}

\author{R.A. Windhorst, S.H. Cohen, A.N. Straughn, R.E. Ryan Jr., 
	N.P. Hathi, R.A. Jansen}
\address{Department of Physics and Astronomy, Arizona State University, Box
	871504, Tempe, AZ 85287;\\Email: Rogier.Windhorst@asu.edu}
\author{A.M. Koekemoer, N. Pirzkal, C. Xu, B. Mobasher, S. Malhotra,
	L. Strolger, J.E. Rhoads}
\address{Space Telescope Science Institute, Baltimore, MD 21218}

\runtitle{Galaxy Assembly and Supermassive Black-Hole Growth}
\runauthor{R. Windhorst}

\maketitle

\begin{abstract}
In this paper, we address whether the growth of supermassive black-holes has
kept pace with the process of galaxy assembly. For this purpose, we first 
searched  the Hubble Ultra Deep Field (HUDF)  for ``tadpole galaxies'', which
have a knot at one end plus an extended tail. They appear dynamically unrelaxed
--- presumably early-stage mergers --- and make up $\sim$6\% of the field
galaxy population. Their redshift distribution follows that of field galaxies,
indicating that --- if tadpole galaxies are indeed dynamically young --- the
process of galaxy assembly generally kept up with the reservoir of field
galaxies as a function of epoch.

\n Next, we present a search for HUDF objects with point-source components that
are optically variable (at the \cge 3.0$\sigma$ level) on timescales of
weeks--months. Among 4644 objects to \iAB$\simeq$28.0 mag (10$\sigma$), 45 have
variable point-like components, which are likely weak AGN. About $\sim$1\% of
all field objects show variability for 0.1\cle z\cle 4.5, and their redshift
distribution is similar to that of field galaxies. Hence supermassive
black-hole growth in weak AGN likely also kept up with the process of galaxy
assembly. However, the faint AGN sample has almost no overlap with the tadpole
sample, which was predicted by recent hydrodynamical numerical simulations. This
suggests that tadpole galaxies are early-stage mergers, which likely preceded
the ``turn-on'' of the AGN component and the onset of visible point-source
variability by \cge 1 Gyr. 
\begin{keyword}
galaxies: mergers \sep galaxies: formation \sep galaxies: active galactic
nuclei \sep Supermassive Black Holes 
\end{keyword}
\end{abstract}

\end{frontmatter}


\section{\vspace*{-4pt}Introduction} \label{introduction}

From the WMAP polarization results (Kogut \etal 2003), population III stars
likely existed at z$\simeq$20. These massive stars (\cge 250\Mo) are expected to
produce a large population of black holes (BH; \Mbh\cge 150\Mo; Madau \& Rees
2001). Since there is now good dynamical evidence for the existence of
supermassive (\Mbh$\simeq$10$^{6}$--10$^{9}$\Mo) black holes (SMBH's) in the
centers of galaxies at z$\simeq$0 (Kormendy \& Richstone 1995; Magorrian,
Tremaine, \& Richstone 1998; Kormendy \& Gebhardt 2001), it is important to
understand how the SMBH's seen at z$\simeq$0 have grown from lower mass BH's
at z$\simeq$20. A comprehensive review of SMBH's is given by Ferrarese \& Ford
(2004). \ One sugges-\vfill\eject

\n tion is that they ``grow'' through repeated mergers of
galaxies which contain less massive BH's, so the byproduct is a larger single
galaxy with a more massive BH in its center. The growth of this BH may then be
observed via its AGN activity. If this scenario is valid, there may be an
observable link between galaxy mergers and increased AGN activity (Silk \& Rees
1998). Therefore, studying this link as a function of redshift could give
insight into the growth of SMBH's and its relation to the process of galaxy
assembly. 

Recent numerical simulations addressed some long-standing issues in the
dissipational collapse scenario by including previously-neglected energetic
feedback from central SMBH's during the merging events (\eg, Robertson \etal
2005). They emphasize the relationship between the central BH mass and the
stellar velocity dispersion, which confirms the link between the growth of
BH's and their host galaxies (di Matteo \etal 2005; Springel \etal 2005ab).
The present study provides observational support for these models at
cosmological redshifts.

\section{The Hubble Ultra Deep Field data} \label{data}

The Hubble Ultra Deep Field (HUDF; Beckwith \etal 2005) is the deepest optical
image of a slice of the Universe ever observed. It consists of 400 orbits with
the \HST\ Advanced Camera for Surveys (ACS) observed over a period of four months
in four optical bands ($BVi'z'$). These are supplemented in the $JH$-bands with
the Near-Infrared Camera and Multi-Object Spectrograph (NICMOS; Bouwens \etal
2004). The HUDF reaches $\sim$1.0 mag deeper in $B$ and $V$ and $\sim$1.5 mag
deeper in \iAB\ and \zAB\ than the equivalent filters in the Hubble Deep Field
(HDF, Williams \etal 1996). 

A large number of galaxies in the HUDF appear dynamically unrelaxed, which
suggests they must play an important role in the overall picture of galaxy
formation and evolution. In particular, we notice many galaxies with a
knot-plus-tail morphology, which constitute a well-defined subset of the
irregular and peculiar objects in the HUDF that is uniquely measurable.
According to di Matteo \etal (2005), this morphology appears to represent an
\emph{early} stage in the merging of two nearly-equal mass galaxies. They are
mostly linear structures, resembling the ``chain'' galaxies first reported by
Cowie, Hu, \& Songaila (1995). When more than two clumps come together, these
objects may be more akin to the luminous diffuse objects and clump clusters
(Conselice \etal 2004; Elmegreen, Elmegreen, \& Sheets 2004; Elmegreen,
Elmegreen, \& Hirst 2004), or other types of irregular objects (Driver \etal
1995; van den Bergh 2002). 

Since the HUDF data was observed over a period of four months, it also provides
a unique opportunity to search for variability in all types of objects to very
faint flux levels, such as faint stars, distant supernovae (SNe), and weak
active galactic nuclei (AGN). From all objects detected in the HUDF, we
therefore selected the subset of tadpole galaxies and variable objects, and
analyzed their properties in the \iAB-band, where the HUDF images are deepest
and have the best temporal spacing over four months. Yan \& Windhorst (2004b)
discuss how the \iAB-selection result in a small bias against objects at z\cge
5.5 in the high redshift tail of the redshift distribution. However, tadpole
galaxies at z$\simeq$5.5 \emph{do} exist (\eg, Rhoads \etal 2005). Since most
HUDF objects have z\cge 1.5, the \iAB-band images sample the rest-frame UV,
where AGN are known to show more variability (Paltani \& Courvoisier 1994). 

To address whether supermassive black-hole growth kept pace with galaxy
assembly, we will present in this paper the redshift distribution of both
tadpole galaxies and weak variable AGN in the HUDF, and compare these with the
redshift distribution of the general field galaxy population. 

\section{Tadpoles as Proxy to Galaxy Assembly} \label{sampleselection}

The steps to select galaxies with the characteristic ``tadpole'' shape are
described in Straughn \etal (2006). In short, objects of interest have a
bright ``knot'' at one end and an extended ``tail'' at the other. Two
different source catalogs were made to \iAB=28.0 mag using \texttt{SExtractor}
(Bertin \& Arnouts 1996): a highly deblended catalog containing many point-like
sources, including the knots of potential tadpole galaxies, and a
low-deblending catalog containing extended sources, including the tadpole's
tails. 

First, the knots of the tadpole galaxies were selected by setting an axis-ratio
limit. ``Knots'' were defined from the highly deblended catalog with an axis
ratio rounder than some critical value ($b/a$$>$0.70). ``Tails'' are elongated
objects selected from the low-deblending catalog with $b/a$$<$0.43. Tadpoles
were defined when a knot was within a certain distance of the geometrical
center of a tail, namely $<$4$a$ (in semi-major axis units of the tail). We
also required that the knot be $>$0.1$a$ from the tail's geometrical center,
since we are searching for asymmetric objects, and want to eliminate upfront as
many of the true edge-on mid-type spiral disks as possible. The tadpole
candidates also must have the knot near one end of the tail, hence we selected
only those tails and knots with a relative position angle
$\Delta\theta$$\leq$20$^{\circ}$, as measured from the semi-major axis of the
tail. This prevented including knots and tails that appear close together on
the image, but are not physically part of the same galaxy. 

Our final sample contains 165 tadpole galaxies, a subset of which is shown in
Fig. 1. These were selected from 2712 objects in the low-deblending HUDF catalog
to \iAB=28.0 mag. Less than 10\% of the selected tadpoles appear as normal
edge-on disk galaxies. Fig. 2 shows a significant overabundance of knots near
the end of the elongated diffuse structures ($\Delta\theta$\cle 10$^{\circ}$)
as compared to randomly distributed knots. Hence, the majority of tadpoles are
not just chance alignments of unrelated knots. Instead, we believe they are
mostly linear structures which are undergoing recent mergers. Their redshift
distribution is shown in Fig. 3a--3b.

\section{Faint Variable Objects as Proxy to SMBH Growth} \label{AGNselection}

Our HUDF variable object study is described in Cohen \etal (2006). Individual
cosmic-ray (CR) clipped images and weight maps were used with {\it
multidrizzle} (Koekemoer \etal 2002) to create four sub-stacks of
approximately equal exposure times that cover 0.4--3.5 months timescales. These
used the same cosmic-ray maps and weight maps as the full-depth HUDF mosaics.
All four epochs were {\it drizzled} onto the {\it same} output pixel scale
(0\arcspt030/pixel) and WCS frame as the original HUDF. Since we are searching
for any signs of variability, we used a liberal amount of object deblending in
the \texttt{SExtractor} catalogs, which used a 1.0$\sigma$ detection threshold
and a minimum of 15 connected pixels (\ie, approximately the PSF area) above sky. This
allows pieces of merging galaxies to be measured separately, to increase the
chance of finding variable events in point-source components. Since each of the
four epoch stacks have half the S/N-ratio of the full HUDF, the sample studied
for variability contains 4644 objects to \iAB\cle 28.0 mag (\cge 10$\sigma$). 

\vspace*{6pt}

The ACS/WFC PSF varies strongly with location on the CCD detectors, and with
time due to orbital ``breathing'' of the \HST\ Optical Telescope Assembly. Hence,
we {\it cannot} use small PSF-sized apertures to search for nuclear
variability, as could be done by Sarajedini \etal (2003a) for the much larger
WFPC2 pixel-size and the {\it on-axis} location of the WFPC2 camera. Instead,
we had to use {\it total} magnitudes of the highly deblended ACS objects. Even
though our total flux apertures may encompass the whole galaxy, any variability
must come from a region less than the 0\arcspt 084 PSF in size, due to the
light-travel time across the variability region. 

\vspace*{6pt}

The four epoch catalogs were compared to each other, resulting in six diagrams
similar to Fig. 4a, which show the change in measured total magnitudes in
matched apertures as a function of the full-depth HUDF flux. The flux-error
distribution was determined iteratively for each pair of observations, such
that 68.3\% of the points lie within the boundaries of the upper and lower
1.0$\sigma$ lines that represent the Gaussian error distribution (Fig. 4a). In
order to demonstrate the Gaussian nature of this error distribution at all flux
levels, the $\Delta mag$-data were divided by the 1.0$\sigma$ model line, and
histograms were computed for the resulting normalized $\Delta mag$ data at
various flux-levels in Fig. 5. These histograms are well fit by normalized
Gaussians with $\sigma$$\simeq$1.0. The HUDF noise distribution is not perfectly
Gaussian, but with 288 independent exposures in the \iAB-band, the error
distribution is as close to Gaussian as seen in any astronomical CCD
application. Once the $\pm$1.0$\sigma$ lines were determined, we find all
objects that are at least 3.0$\sigma$ outliers. Most outliers in Fig. 5 at
$\Delta mag$\cge 3.0$\sigma$ are due to object variability, after pruning large
objects without visible point sources which suffered from \texttt{SExtractor}
deblending errors. In Fig. 4a, we show the $\pm$1$\sigma$, $\pm$3$\sigma$, and
$\pm$5$\sigma$ lines, along with the actual data. The choice of 3.0$\sigma$
implies that we should expect 0.27\% random contaminants. 

In total, we find 45 out of 4644 objects that show the signatures of AGN
variability. These are variable at the \cge 3.0$\sigma$ level, have a compact
region indicative of a point source, and are devoid of visible image defects or
object splitting issues. Less than one of these 45 is expected to be a random
contaminant.  In total, 577 candidates were rejected due to crowding or
splitting issues, or due to the lack of a visible point source.  Fig. 4b shows
the number of $\sigma$ by which each object varied for each of the 6 possible
epoch-pairs. The colored symbols are for the 45 ``best'' candidates. Another
57 objects were found that are ``potentially'' variable candidates. The 
four-epoch light-curves for these 45 variable candidates are shown in Fig. 6.
Of these, 49\% were discovered from a single epoch-pair (usually indicative of
a global rise or decline as a function of time in the light-curve), 43\% in two
epoch-pairs, and only 5\% (2 objects) in 3 epoch-pairs. Further details are
given in Cohen \etal (2006). In summary, the variability fraction on a
timescale of few months (rest-frame timescale few weeks to a month) is at
least 1\% of all HUDF field galaxies. 

Since the HUDF is in the Chandra Deep Field--South (CDF-S, Rosati \etal 2002), 
there exists deep X-ray data. Within the HUDF, there are 16 Chandra sources
(Koekemoer \etal 2004,2006), and we detect four of these as variable in the
optical. One of these is a mid-type spiral with \iAB$\simeq$21.24 mag, that
belongs to a small group of interacting galaxies. Two others are optical point
sources with \iAB$\simeq$21.12 mag and $\simeq$24.79 mag, showing little or no
visible host galaxy. Both have measured spectroscopic AGN emission-line
redshifts at z$\simeq$3 (Pirzkal \etal 2004). The detection of 25\% of the
Chandra sources as optically variable in the HUDF data shows that the
variability method employed here is a reliable way of finding the AGN that are
not heavily obscured. 

The faint object variability in the HUDF is most likely due to weak AGN, given
the timescales and distances involved. Strolger \& Riess (2005) found only one
moderate redshift SN in the HUDF, so SNe cannot be a significant source of
contamination in our sample. Several other possible source of incompleteness in
the variability study must be addressed. Non-variable AGN, or AGN that only
vary on timescales much longer than 4 months, or optically obscured AGN would
not have been detected with our UV--optical variability method. Sarajedini
\etal (2003ab) had two HDF epochs 5--7 years apart, and found 2\% of the HDF
to be variable. It is thus possible that our sampled times-scale shorter than 4
months missed a factor \cge 2 of all AGN --- the ones variable on longer
time-scales. 

\section{The Redshift Distribution of Tadpole Galaxies and Faint Variable
Objects} \label{redshiftdistribution}

We calculate photometric redshifts of all HUDF galaxies to \iAB=28.0 mag (\cge
20$\sigma$) from the $BVi'z'$ photometry using \texttt{HyperZ} (Bolzonella \etal 2000),
plus NICMOS $JH$ (Thompson \etal 2005) and VLT ISAAC $K$-band images where
available. When compared to published spectroscopic redshifts for 70 CDF-S
objects (Le F{\' e}vre \etal 2005), our photometric redshifts have an rms
scatter of 0.15 in $\delta$=(z$_{phot}$--z$_{spec}$)/(1+z$_{spec}$) if all 70
objects are included, and 0.10 in $\delta$ when we reject a few of the most
obvious outliers. 

The redshift distribution of all HUDF galaxies (solid line in Fig. 3a) is as
expected, with the primary peak at z\cle 1.0 and a generally declining tail at
z$\simeq$4--5. These trends were also seen in the HDF field galaxies (Driver
\etal 1998). A deficit of objects is apparent at z$\simeq$1--2 due to the lack
of UV spectral features crossing the $BViz$(+$JH$) filters. Unlike the HDF,
this deficit occurs because the HUDF does not yet have
deep enough F300W or $U$-band data. The resulting redshift bias, however, is
the \emph{same} for both tadpoles, variable objects and the field galaxy
population, and so divides out in the subsequent discussion. Within the
statistical uncertainties in Fig. 3a, the shape of the tadpole galaxy redshift
distribution follows that of the field galaxies quite closely. This suggests
that if tadpole galaxies are indeed dynamically young objects related to
early-stage mergers, they seem to occur in the same proportion to the field
galaxy population at all redshifts. Tadpole galaxies may therefore be good
tracers of the galaxy assembly process. The ratio of the two redshift
distributions N(z) and the resulting percentage of tadpole galaxies is plotted
in Fig. 3b. Overall, the percentage of tadpole galaxies is roughly constant at
$\sim$6\% with redshift to within the statistical errors for the redshifts
sampled (0.1\cle z\cle4.5).

In Fig. 7a, we show the photometric redshift distribution for all HUDF objects
with \iAB\cle 28.0 mag, and for our best 45 variable candidates. Their 
redshift distribution follows that of the field galaxies in general, \ie,
there is no redshift where faint object variability was most prevalent. We plot
in Fig. 7b the ratio of the N(z) for variable objects to that of field
galaxies, and show that the weak variable AGN fraction is roughly constant at
approximately 1\% over all redshifts probed in this study. 

\section{Discussion and Conclusions}

The fact that about 6\% of all field galaxies are seen in the tadpole stage 
is an important constraint to hierarchical simulations. Springel \etal
(2005ab) predict a tadpole-like stage $\sim$0.7--1.5 Gyr after a major merger
begins, suggesting that the tadpole morphology represents an early-merger stage
of two galaxies of roughly comparable mass. If this 6\% indicates the fraction
of time that an average galaxy in the HUDF spends in an early-merger stage
during its lifetime, then every galaxy may be seen  temporarily in a tadpole
stage for $\sim$0.8 Gyr of its lifetime,  and may have undergone $\sim$10-30
mergers during its lifetime (Straughn \etal 2006).  More complex mergers
involving multiple components may lead to irregular/peculiar and train-wreck
type objects, and the luminous diffuse objects or clump-clusters, which
dominate the galaxy counts at faint magnitudes (Driver \etal 1998). Given that
tadpoles only trace a certain type and stage of merging galaxies, the above
statistics are a lower limit on the number of all mergers. 

The question arises if tadpole galaxies and objects with point-sources that
show signs of variability are drawn from the same population. Among our 165
tadpole galaxies, none coincide with the sample of 45 variable objects or with
the CDF-S X-ray sources S (Alexander \etal 2005). At most one or two of the
variable candidates resemble the tadpole galaxies of Straughn \etal (2006). 

A factor of three of all AGN may have been missed, since their UV--optical flux
was obscured by a dust-torus. In the AGN unification picture, AGN cones are
two-sided and their axes are randomly distributed in the sky, so that an
average cone opening-angle of $\omega$ implies that a fraction
1--sin($\omega$) of all AGN will point in our direction. If
$\omega$$\simeq$45$^\circ$ (\eg, Barthel 1989), then every optically detected
AGN (QSO) represents 3--4 other bulge-dominated galaxies, whose AGN reflection
cone didn't shine in our direction. Hence, their AGN may remain obscured by the
dust-torus. Such objects could be visible to Chandra in X-rays or to Spitzer
in the mid-IR, although the available Chandra and Spitzer data are not deep
enough to detect all HUDF objects to AB$\simeq$28 mag. 

Together with the factor of \cge 2 incompleteness in the HUDF variability
sample due to the limited time-baseline sampled thus far, the actual fraction
of weak AGN present in these dynamically young galaxies may thus be a factor of
$\gtrsim$6--8$\times$ larger than the 1\% variable AGN fraction that we found 
in the HUDF. Hence, perhaps as many as $\gtrsim$6--8\% of all field galaxies
may host weak AGN, only $\sim$1\% of which we found here, and another \cge 1\%
could have been found if longer time-baseline had been available. Another
factor of 3--4 of AGN are likely missing because they are optically obscured,
The next generation of X-ray and IR telescopes (Windhorst \etal 2006ab) and
longer optical time-baselines are needed to detect all weak AGN in the HUDF.

Recent state-of-the-art hydrodynamical models (di Matteo \etal 2005; 
Springel \etal 2005ab; Hopkins \etal 2005) suggest that during (major)
mergers, the BH accretion rate peaks considerably {\it after} the merger
started, and {\it after} the star-formation rate (SFR) has peaked. Their models
suggest that, for massive galaxies, a tadpole stage is seen typically about 0.7
Gyr after the merger started, but $\sim$1 Gyr before the SMBH accretes most of
its mass, which is when the galaxy displays strong visible AGN activity. Since
the lifetimes of QSO's and radio-galaxies are known to be \cle
(few$\times$10$^7$)--10$^8$ years (Martini 2004; Grazian \etal 2004, Jakobsen
\etal 2003), these models thus imply that the AGN stage is expected to occur
considerably (\ie, \cge 1--1.5 Gyr) \emph{after} the early-merger event during
which the galaxy is seen in the tadpole stage. 

The observed lack of overlap between the HUDF tadpole sample and the weak
variable AGN sample thus provides observational support for this prediction.
Hopkins \etal (2005) have quantified the timescales that quasars will be
visible during merging events, noting that for a large fraction of the
accretion time, the quasar is heavily obscured. In particular, their
simulations show that during an early merging phase --- our observed tadpole
phase --- the intrinsic quasar luminosity peaks, but is completely optically 
obscured. Only after feedback from the central quasar clears out the gas, will
the object become visible as an AGN. This should be observable by Spitzer in
the mid-IR as a correspondingly larger fraction of IR-selected obscured faint
QSO's. To study the relation between galaxy assembly and SMBH growth in detail,
we need deeper surveys at longer wavelengths with the James Webb Space
Telescope (JWST; Windhorst \etal 2006a). The JWST photometric {\it and} PSF
stability are crucial for this, since many of our HUDF objects show significant
variability of less than a few percent in total flux. 

This research was partially funded by NASA grants GO-9793.08-A and
AR-10298.01-A, NASA JWST grant NAG5-12460, the NASA Space Grant program at
ASU, and the Harriet G. Jenkins Predoctoral Fellowship Program.


\ve
\n\makebox[\textwidth][s]{
   \psfig{file=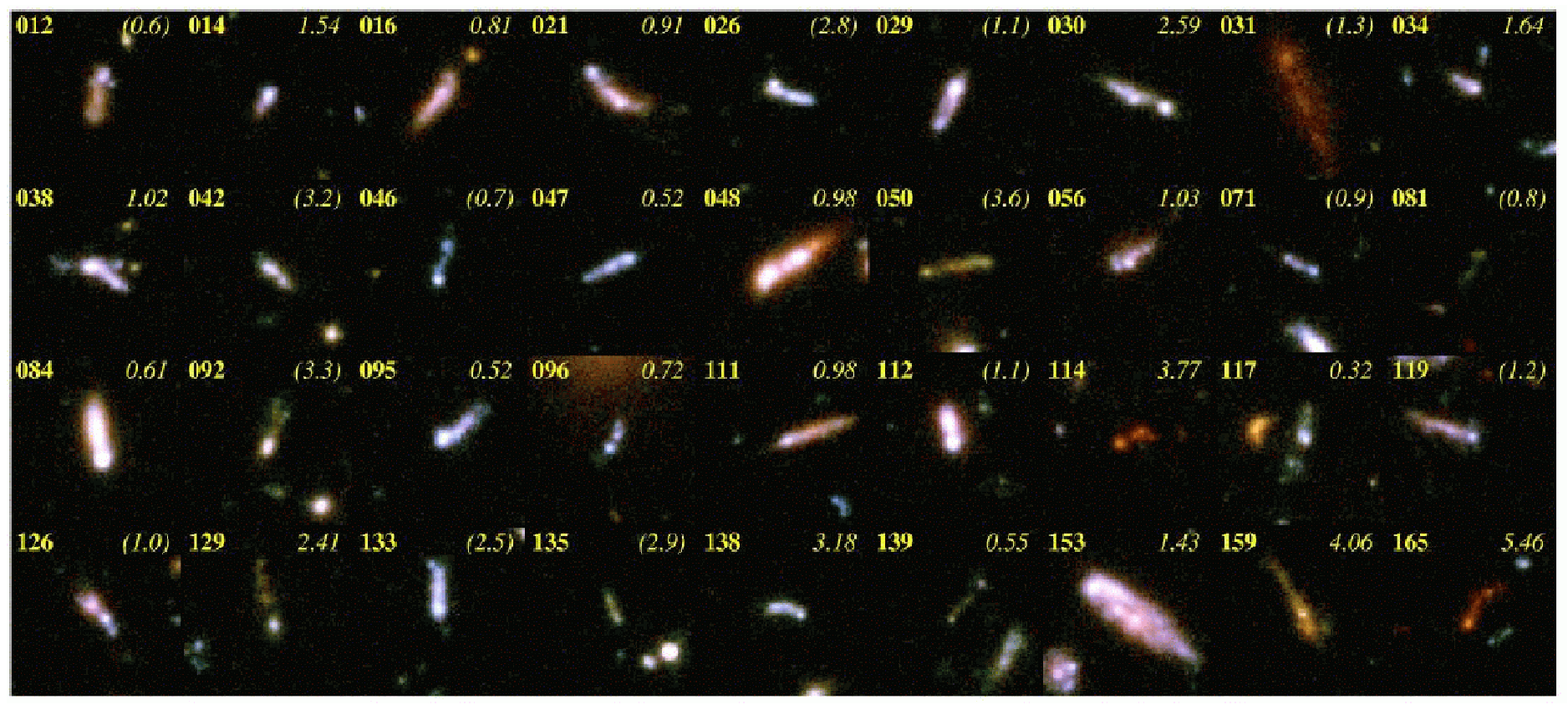,width=0.99\textwidth,angle=0}
}

\sn \begin{minipage}{1.00\textwidth}
{\footnotesize\baselineskip=10pt {\bf Fig. 1.}\ \iAB-band mosaic of a
subset of the HUDF tadpole galaxy sample. Stamps are 3 arcsec on a side. The
vast majority of our tadpole sample contains the distinctive knot-plus-tail
morphology.}
\end{minipage}

\vspace*{0.5cm}

\sn\makebox[\textwidth][s]{
   \psfig{file=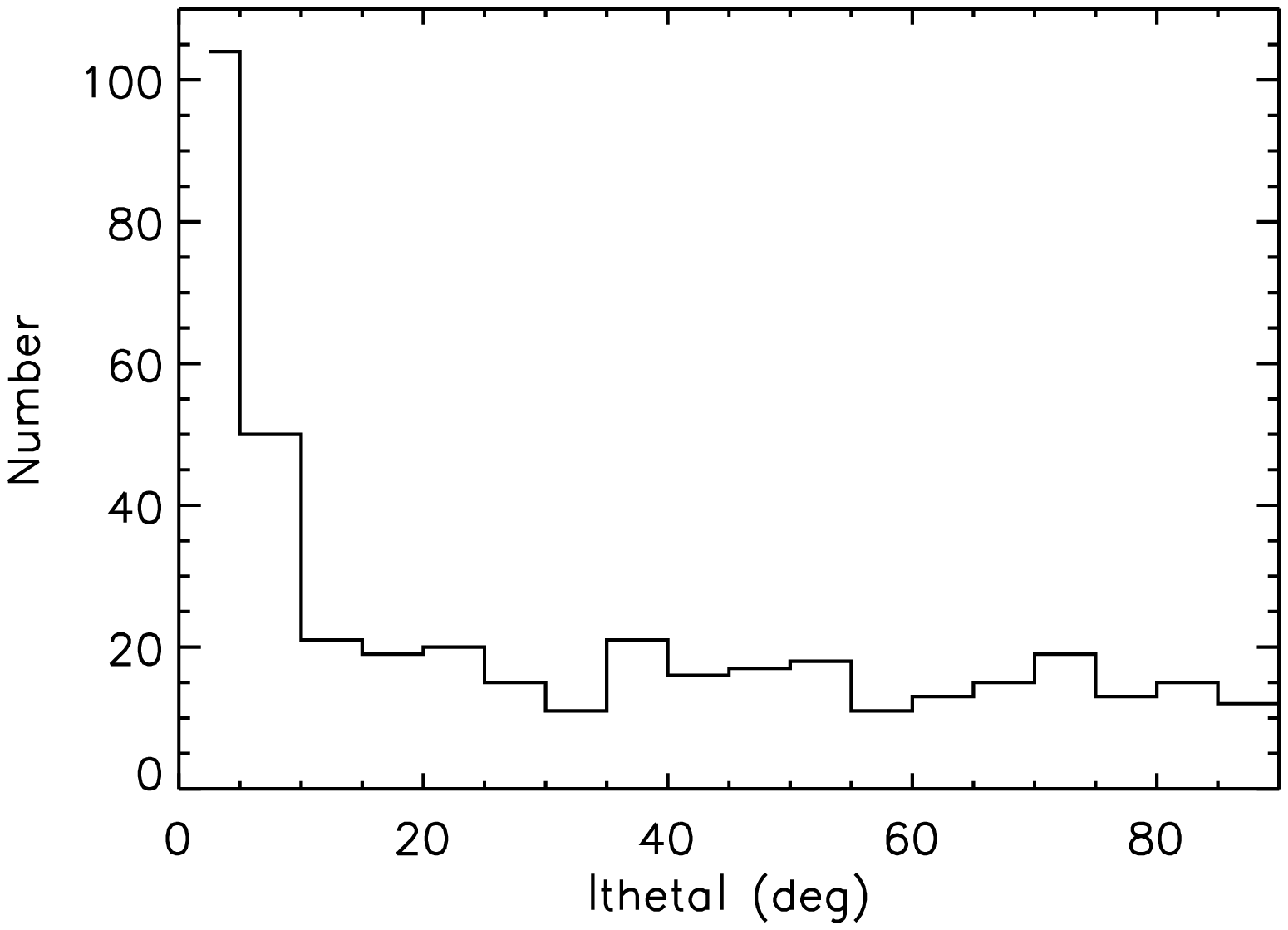,width=0.45\textwidth,angle=0}
}

\sn \begin{minipage}{1.00\textwidth}
{\footnotesize\baselineskip=10pt {\bf Fig. 2.}\ Distribution of position
angle $\theta$, measured from the major axis of the diffuse component, for
all off-centered knots found within r$\leq$4$a$ (\cle 2'') from the center
of an elongated diffuse object in the HUDF. There is a clear excess of
knots near $\vert\Delta\theta\vert$$\simeq$0$^\circ$, confirming the linear
structure of most tadpoles.}
\end{minipage}

\clearpage

\n\makebox[\textwidth]{
   \psfig{file=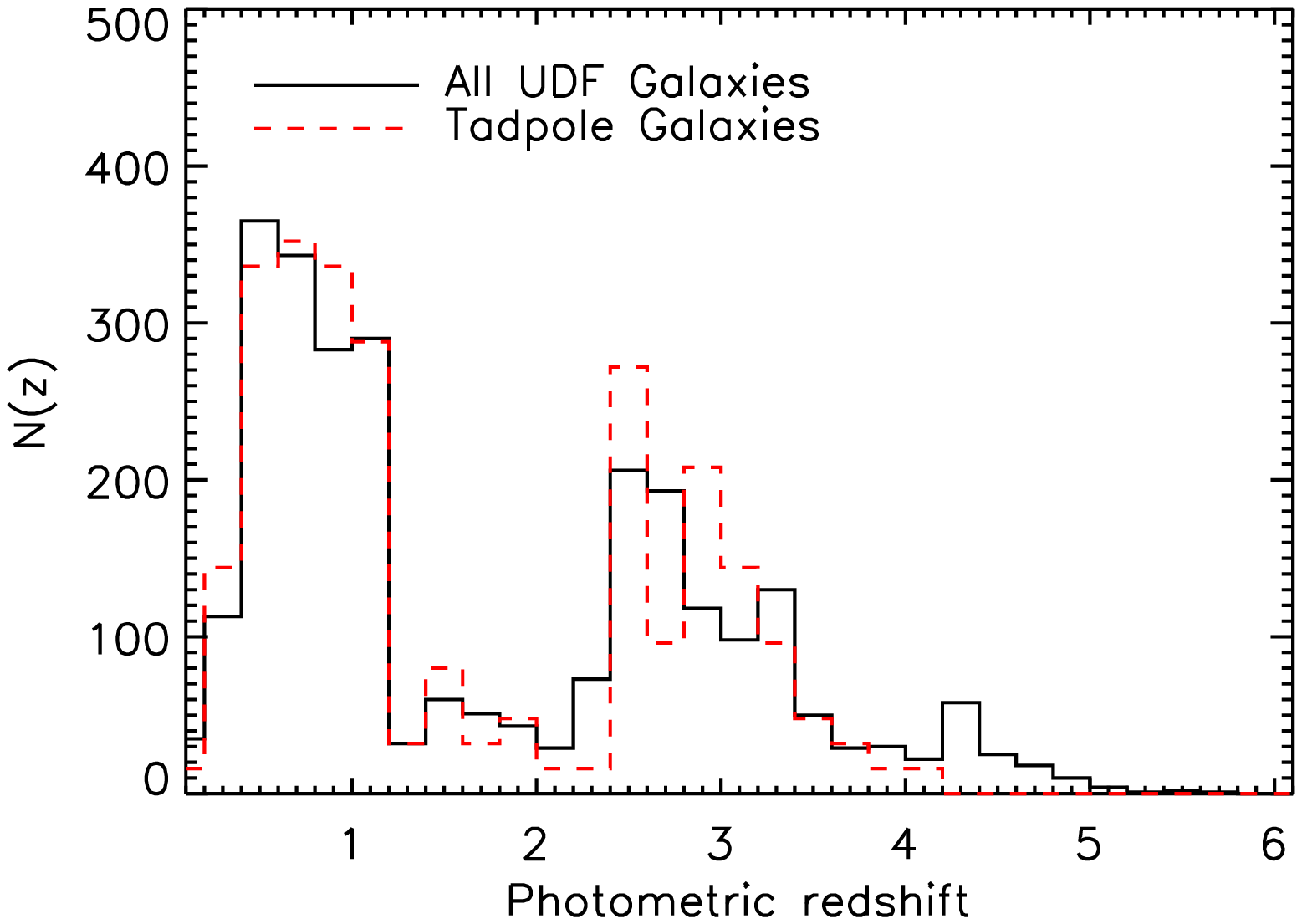,width=0.50\textwidth,angle=0} 
   \psfig{file=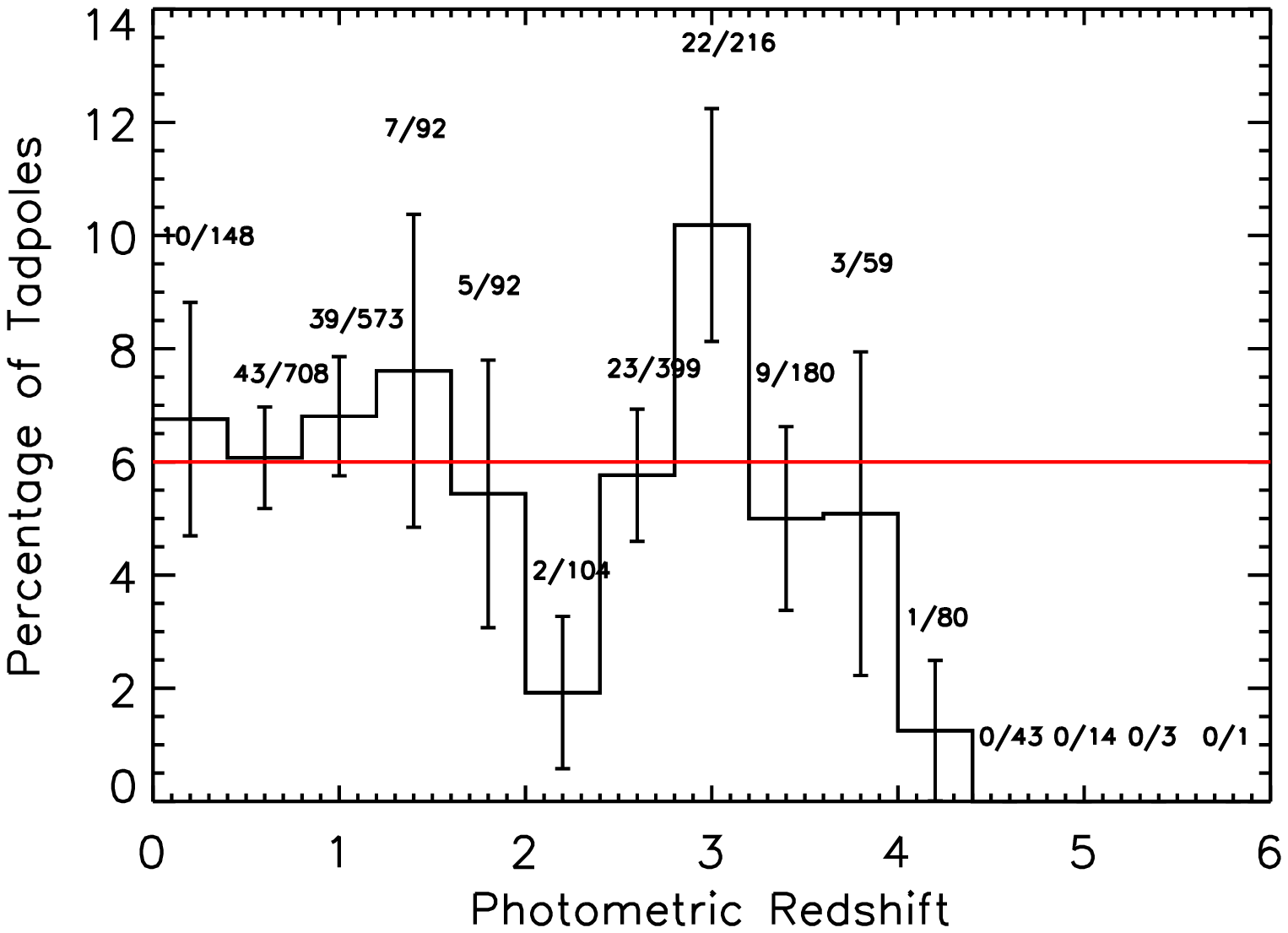,width=0.50\textwidth,angle=0}
}
\sn \begin{minipage}{1.00\textwidth}
\sn {\footnotesize\baselineskip=10pt {\bf Fig. 3a (LEFT).}\ Photometric
redshift distribution of galaxies in the HUDF. The solid black histogram shows
the redshift distribution of all HUDF field galaxies to \iAB=28.0 mag, while the
dashed red histogram shows the redshift distribution of the tadpole galaxies, 
multiplied by 16 for best comparison. }
{\footnotesize\baselineskip=10pt {\bf Fig. 3b (RIGHT).}\ Percentage of total
galaxies that are tadpoles vs. photometric redshift. Within the statistical
errors, $\sim$6\% of all galaxies are seen as tadpoles at all redshifts. }
\end{minipage}



\vspace*{0.5cm}

\sn\makebox[\textwidth]{
   \psfig{file=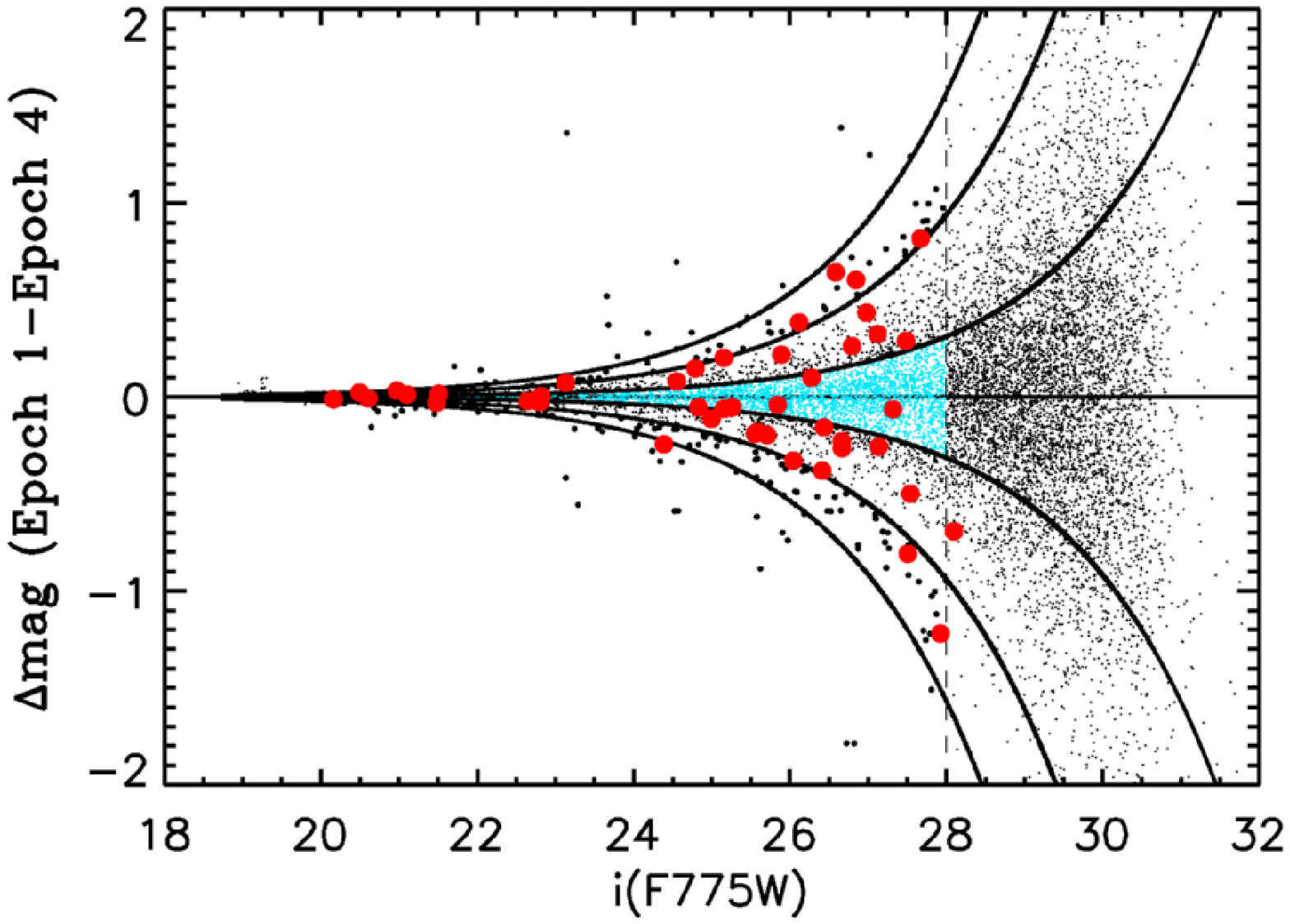,width=0.50\textwidth,angle=0}
   \psfig{file=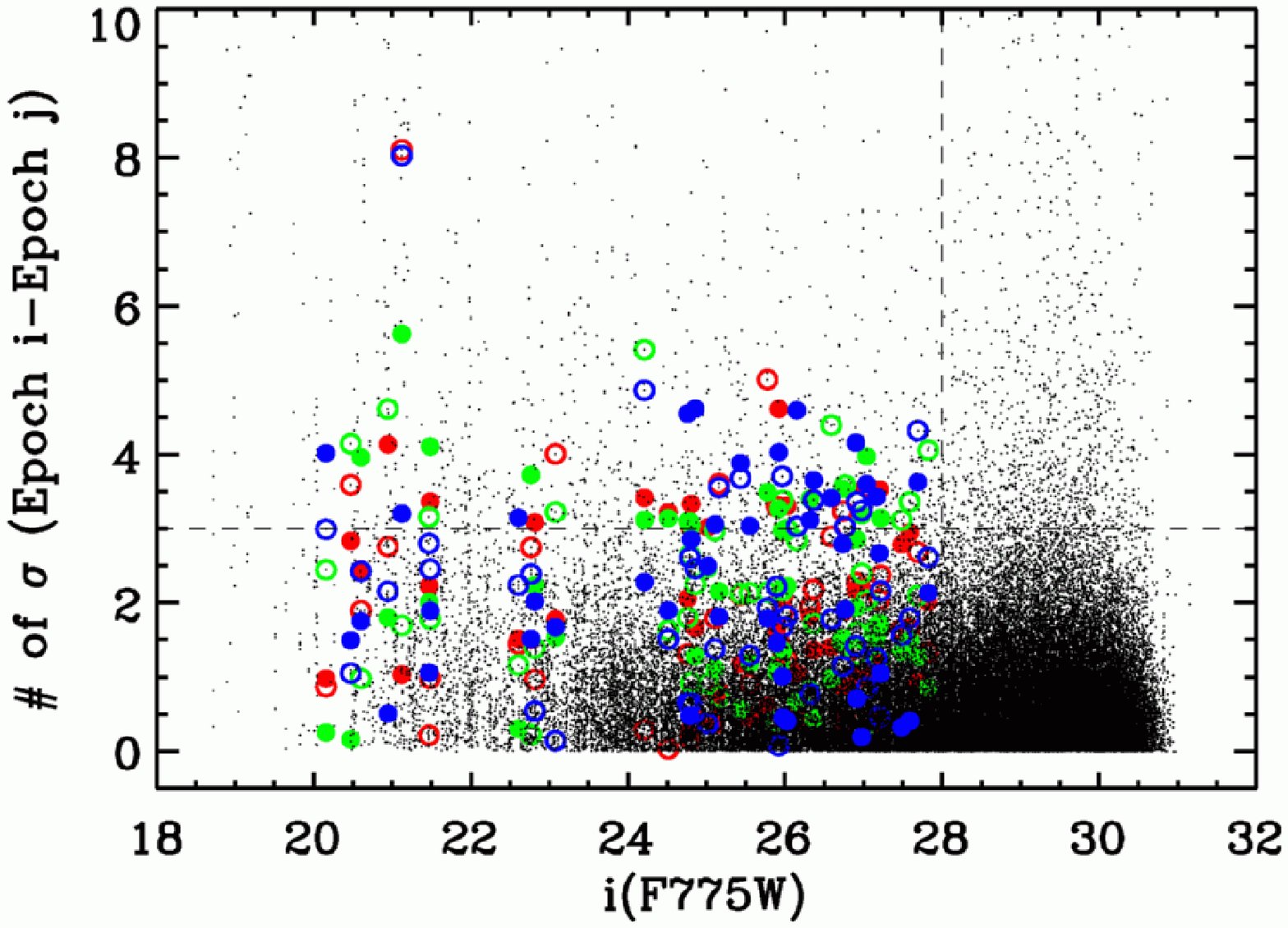,width=0.50\textwidth,angle=0}
}
\sn \begin{minipage}{1.00\textwidth}
\sn {\footnotesize\baselineskip=10pt {\bf Fig. 4a (LEFT).}\ Magnitude
difference between two HUDF epochs of all objects vs. \iAB-band flux from
matched total apertures. The $\pm$1$\sigma$, $\pm$3$\sigma$, $\pm$5$\sigma$
lines are shown. Blue points show the $\vert\Delta mag\vert$ \cle
$\pm$1.0$\sigma$ points used to normalize the error distribution. Large red
points show the ``best''  45 variable candidates from all six possible epoch
combinations, many of which were seen at \cge 3.0$\sigma$ in two or more epoch
combinations. }
\n {\footnotesize\baselineskip=10pt {\bf Fig. 4b (RIGHT).}\ Number of $\sigma$
that each object varies for all six possible epoch combinations. Colored
symbols indicate the ``best'' sample of 45 variable candidates from Fig. 4a,
that are unaffected by local image deblending issues or weight map structures.
Each object appears six times in this plot. }
\end{minipage}

\ve 

\n\ 

\ve
\vspace*{-1.8cm}\hspace*{-0.4cm}
\makebox[0.480\textwidth]{
   \psfig{file=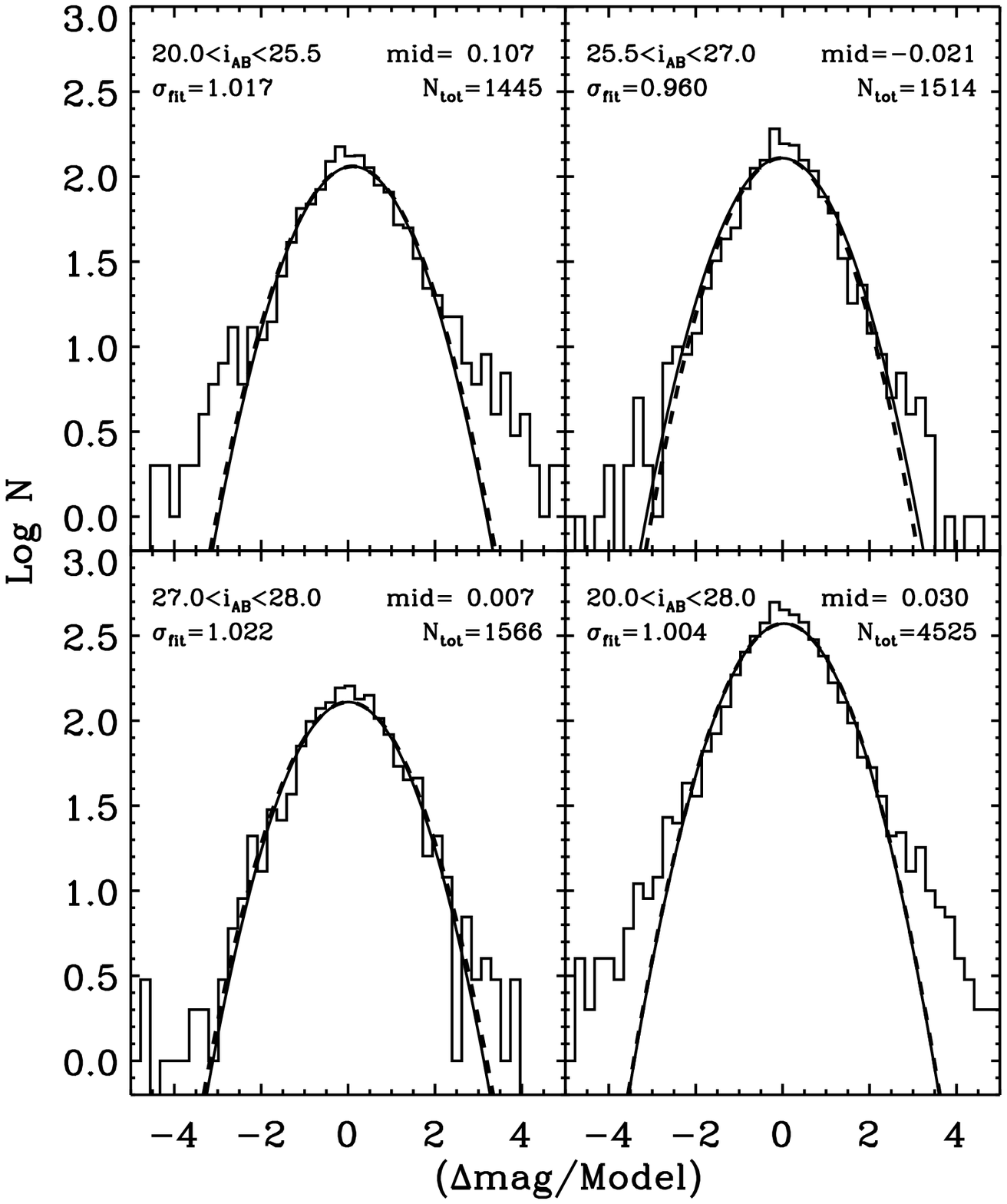,height=0.525\textheight,angle=0}
}\makebox[0.480\textwidth]{
   \psfig{file=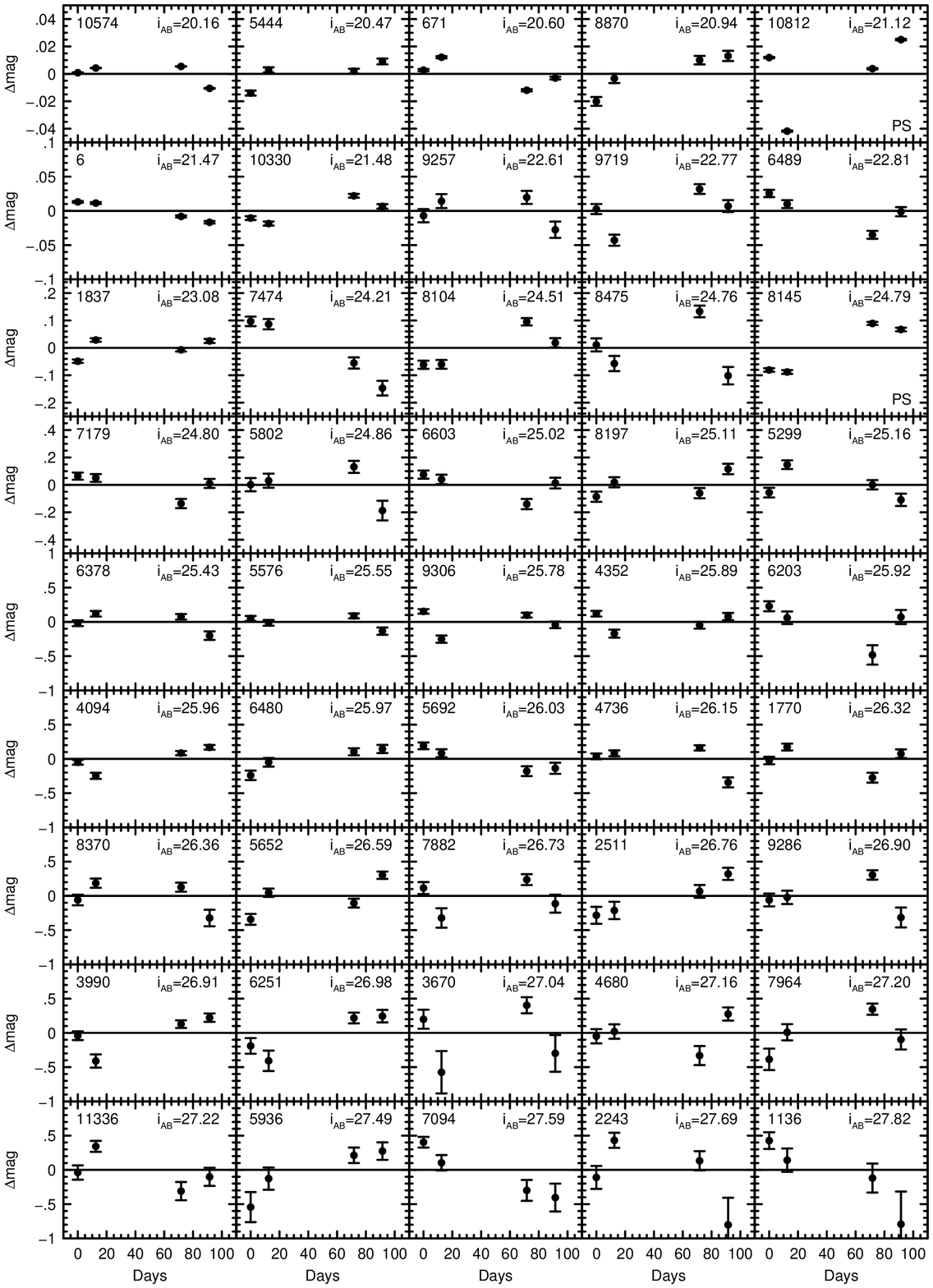,height=0.525\textheight,angle=0}
}

\vspace*{-1.5cm}
\sn\makebox[\textwidth]{
\begin{minipage}[b]{0.48\textwidth}
{\footnotesize\baselineskip=10pt {\bf Fig. 5.}\  Gaussian nature of the
HUDF total-flux error distribution at all flux levels. The $\Delta mag$
data from Fig.~4a were divided by the best-fit model 1.0$\sigma$ lines.
Histograms for the indicated magnitude ranges are well fit by normalized
Gaussians (parabolas in log space) with $\sigma$$\simeq$1.0.  The almost
indistinguishable dashed and solid lines are for the best-fit $\sigma$ 
(indicated in the legends) and for assumed $\sigma$$\equiv$1 Gaussians, 
respectively. Hence, Fig. 5 shows that 3.0$\sigma$ really means 
3.0$\pm$0.1$\sigma$.  All objects with $\Delta mag$\cge 3.0$\sigma$ not
affected by deblending errors are variable candidates. }
\end{minipage}\hspace*{0.04\textwidth}
\begin{minipage}[b]{0.48\textwidth}
{\footnotesize\baselineskip=10pt {\bf Fig. 6.}\  Light curves of the 45 best
candidates with signs of optical point-source variability.  The change in
measured total flux (average$-$individual epoch) is plotted vertically, and
the number of days since the first epoch is plotted horizontally.}
\end{minipage}
}

\n\makebox[\textwidth]{
   \psfig{file=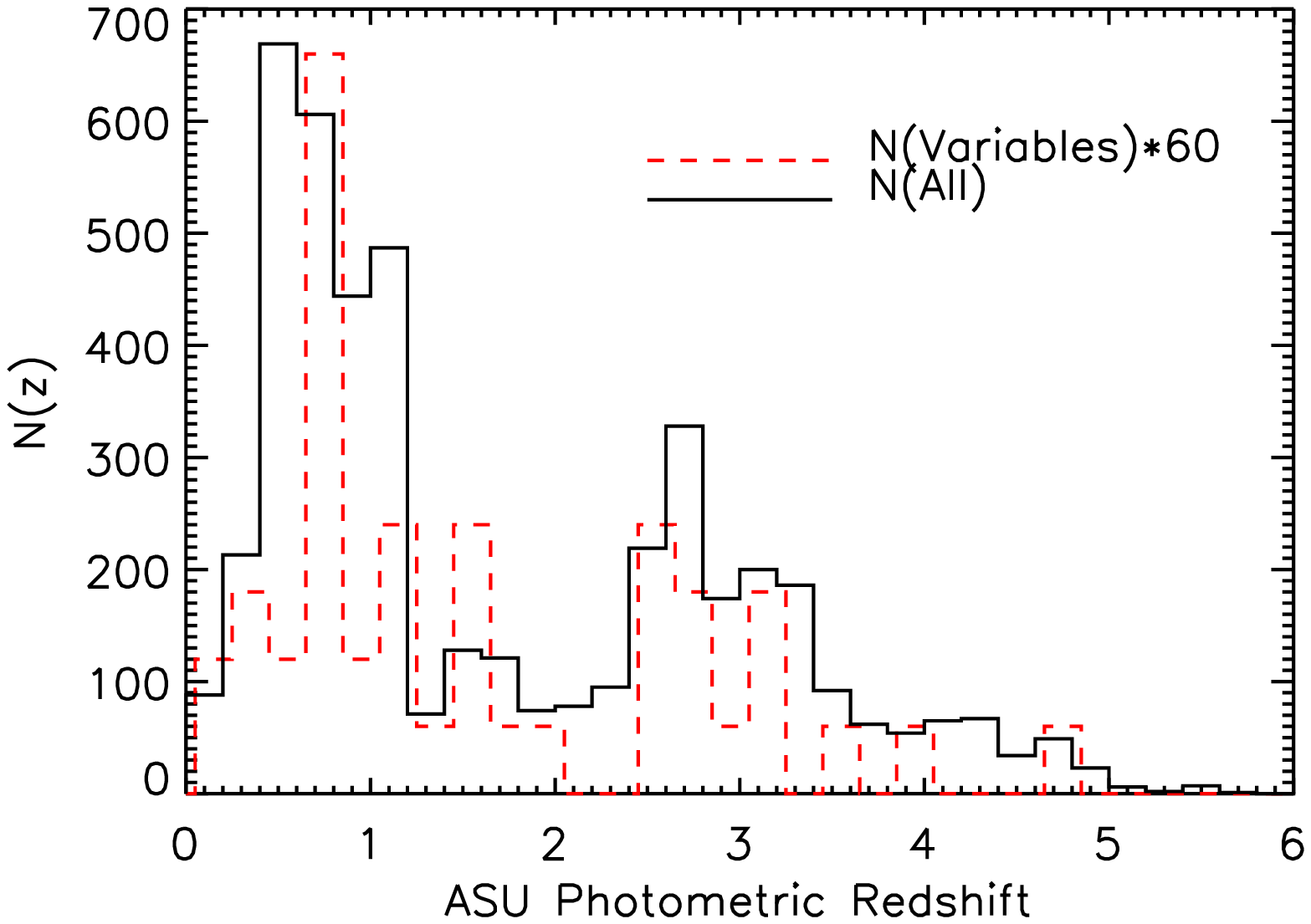,width=0.50\textwidth,angle=0} 
   \psfig{file=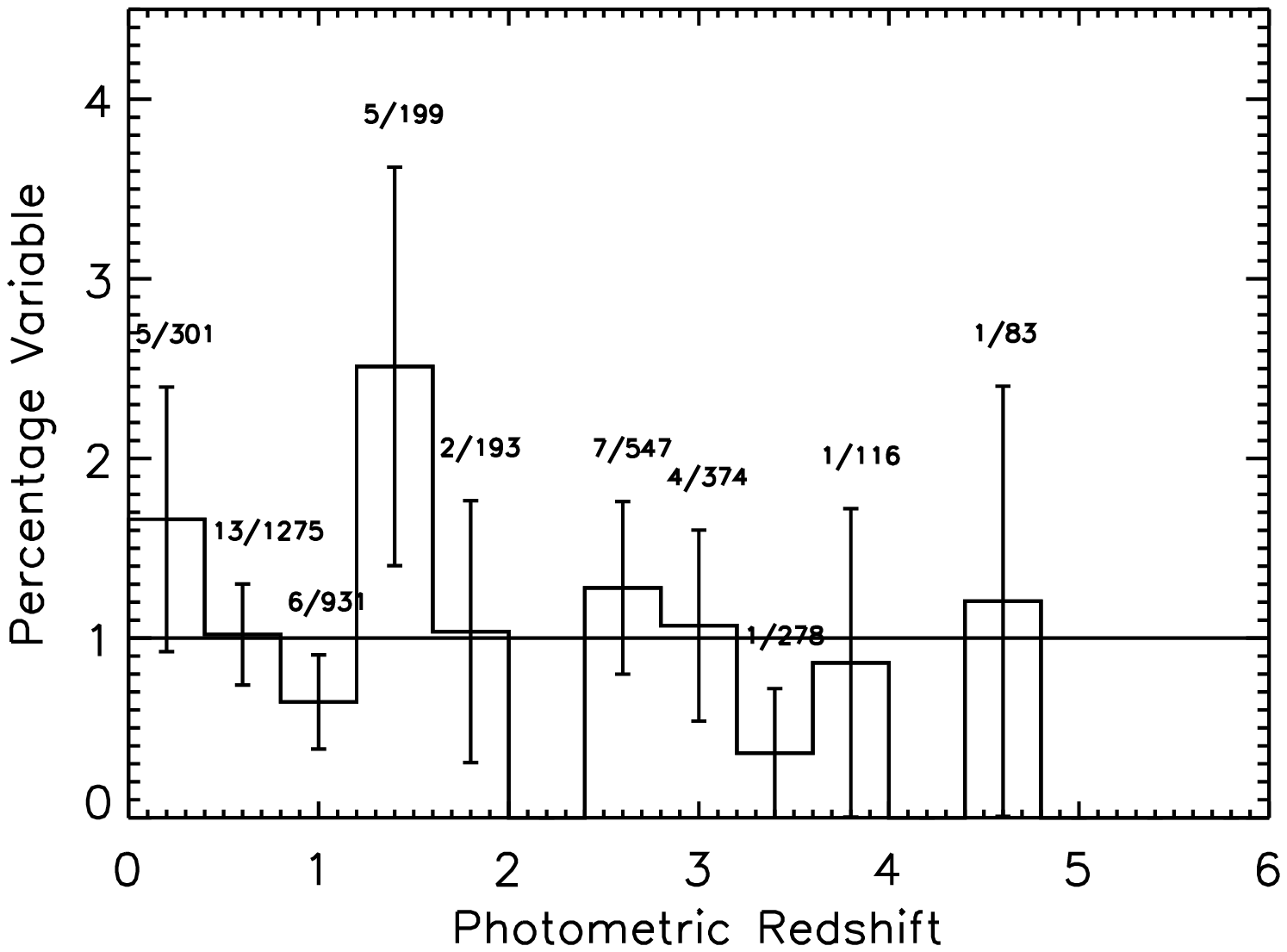,width=0.50\textwidth,angle=0}
}
\sn \begin{minipage}{1.00\textwidth}
\sn {\footnotesize\baselineskip=10pt {\bf Fig. 7a (LEFT).}\ Photometric redshift
distribution of all HUDF field galaxies to \iAB\cle 28.0 mag (solid line), 
and for the ``best'' variable candidates (red dashed line) multiplied by
60$\times$ for best comparison. The redshift distribution of the variable
objects follows that of field galaxies in general.} \n
{\footnotesize\baselineskip=10pt {\bf Fig. 7b (RIGHT).}\ Percentage of HUDF
objects to \iAB \cle 28.0 AB-mag showing variable point sources as a function
of redshift. Within the statistical uncertainties, about 1\% of all HUDF
galaxies show point source variability over the redshift range surveyed (0\cle
z\cle 5). Hence, SMBH growth as traced by the weak AGN fraction keeps pace with
galaxy assembly as traced by the tadpoles in Fig. 3b. }
\end{minipage}


\begin{thebibliography}
\frenchspacing\small\baselineskip=10pt
\bibitem{ax03} Alexander, D.M. \etal 2003, AJ 126, 539
\bibitem{barthel04} Barthel, P.D. 1994, ApJ 336, 606
\bibitem{beck05} Beckwith, S., \etal 2005, AJ, submitted
\bibitem{sex} Bertin, E. \& Arnouts, S. 1996, A\&AS 117, 363
\bibitem{hyper} Bolzonella, M., \etal 2000, A\&A 363,476 
\bibitem{bouwens04} Bouwens, R.J., \etal 2004, ApJ 616, L79
\bibitem{bc93} Bruzual, G. \& Charlot, S. 1993 ApJ 405, 538
\bibitem{coh06} Cohen, S.H., \etal 2006, ApJ 639, in press\\(astro-ph/0511414) 
\bibitem{con04} Conselice, C. \etal 2004, ApJ 600, L139
\bibitem{cow95} Cowie, L. \etal 1995, AJ 110, 1576
\bibitem{dimatteo05} di Matteo, T., \etal 2005, Nature 433, 604 
\bibitem{dri95} Driver, S.P., \etal 1995, ApJ 449, L23 
\bibitem{dri98} Driver, S.P., \etal 1998, ApJ 496, L93--L97 
\bibitem{elm04a} Elmegreen, D., \etal 2004, ApJ 604, L21 
\bibitem{elm04b} Elmegreen, D., \etal 2004, ApJ 603, 74 
\bibitem{ff04} Ferrarese, L. \& Ford, H. 2004, Space Science Rev. 116, 523
\bibitem{gra04} Grazian, A., \etal 2004, AJ 127, 592
\bibitem{hey04} Heymans, C., \etal 2005, MNRAS 361, 160
\bibitem{hop05} Hopkins, P., \etal 2005, ApJ 625, L71 
\bibitem{jak03} Jakobsen, P., \etal 2003, A\&A 397, 891 
\bibitem{driz1} Koekemoer, A.M., \etal 2002, in: ``The 2002 HST Calibration Workshop", Eds.\ S.\ Arribas \etal (Baltimore: STScI), 337 
\bibitem{koe04} Koekemoer, A.M., \etal 2004, ApJ 600, L123
\bibitem{driz2} Koekemoer, A.M., \etal 2006, in preparation
\bibitem{kogut} Kogut, A., \etal 2003, ApJS 148, 161
\bibitem{kor95} Kormendy, J. \& Richstone, D. 1995, ARA\&A 33, 581
\bibitem{kor01} Kormendy, J. \& Gebhardt, K. 2001, in ``20th Texas Symposium'', AIP Conf. Proc. 586, 363 
\bibitem{fev05} Le F{\' e}vre, O., \etal 2005, A\&A 439, 845 
\bibitem{madrees} Madau, P., \& Rees, M. 2001, ApJ 551, L27
\bibitem{mag98} Magorrian, J., \etal 1998, AJ 115, 2285 
\bibitem{mar04} Martini, P. 2004, in ``Co-evolution of Black Holes and Galaxies'' (Cambridge: Cambridge Univ. Press), 170 
\bibitem{palt94} Paltani, S., \& Courvoisier, T. J.-L. 1994, A\&A 291, 74 
\bibitem{grapes} Pirzkal, N., \etal 2004, ApJS 154, 501
\bibitem{rhoads05} Rhoads, J.E., \etal 2005, ApJ 621, 582
\bibitem{rob05} Robertson, B. \etal 2005, (astro-ph/0503369)
\bibitem{ros02} Rosati, P., \etal 2002, ApJ 566, 667
\bibitem{sara1} Sarajedini, V.L., \etal 2003a, ApJ 599, 173 
\bibitem{sara2} Sarajedini, V.L. 2003b, MmSAI 74, 957
\bibitem{silkrees} Silk, J., \& Rees, M. 1998, A\&A 331, L1
\bibitem{spring05} Springel, V., \etal 2005a, ApJ 620, 79 
\bibitem{spring2} Springel, V., \etal 2005b, MNRAS 361, 776 
\bibitem{strau06} Straughn, A.N., \etal 2006, ApJ 639, in press\\(astro-ph/0511423) 
\bibitem{strol05} Strolger, L.-G., \& Riess, A. G. 2005, astro-ph/0503093
\bibitem{rtleg} Thompson, R.I., \etal 2005, AJ 130, 1
\bibitem{vdBergh02} van den Bergh, S. 2002, PASP 114, 797
\bibitem{white78} White, S.D.M. \& Rees, M.J. 1978, MNRAS 183, 341
\bibitem{williams96} Williams, R., \etal 1996, AJ 112, 1335 
\bibitem{wind06} Windhorst, R.A., \etal 2006ab, in ``First Light and Reionization'', Eds. A. Cooray \& E. Barton, New Astron. Rev., in press (astro-ph/0506253) 
\bibitem{yw04b} Yan, H., \& Windhorst, R.A. 2004, ApJ 612, L93
\end{thebibliography}
\end{document}